\documentclass[prd, superscriptaddress,nofootinbib]{revtex4}
\usepackage{graphicx}% Include figure files
\usepackage{dcolumn}% Align table columns on decimal point
\usepackage{slashbox,multirow}
\usepackage{amsmath,amsthm,amssymb}
\begin{document}
%

%\preprint{APS/123-QED}

\title{Late-time tails of a self-gravitating massless scalar field, revisited}

\author{Piotr Bizo\'n}
\affiliation{M. Smoluchowski Institute of Physics, Jagiellonian
University, Krak\'ow, Poland}
\author{Tadeusz Chmaj}
\affiliation{H. Niewodniczanski Institute of Nuclear
   Physics, Polish Academy of Sciences,  Krak\'ow, Poland}
   \affiliation{Cracow University of Technology, Krak\'ow,
    Poland}
\author{Andrzej Rostworowski}
\affiliation{M. Smoluchowski Institute of Physics, Jagiellonian
University, Krak\'ow, Poland}

\date{\today}
\begin{abstract}
We discuss the nonlinear origin of the power-law tail in the long-time evolution of a spherically
symmetric self-gravitating massless scalar field in even-dimensional spacetimes. Using
third-order perturbation method, we derive explicit expressions for the tail (the decay rate and
the amplitude) for solutions starting from small initial data and we verify this prediction via
numerical integration of the Einstein-scalar field equations in four and six dimensions. Our
results show that the coincidence of decay rates of linear and nonlinear tails
 in four dimensions (which has misguided some tail hunters in the past) is in a sense
  accidental and does not hold in higher dimensions.
\end{abstract}

\maketitle

\section{Introduction}
This paper is concerned with the long-time behavior of a spherically symmetric self-gravitating
massless scalar field. This toy-model of gravitational collapse has been intensively studied in
the past leading to valuable insights about the validity of the weak cosmic censorship and
no-hair conjectures. In particular, Christodoulou  proved that there are two generic endstates of
evolution: Minkowski spacetime  for small initial data \cite{christodoulou1} and Schwarzschild
black hole for large initial data \cite{christodoulou2}. In both cases the upper bound for the
rate of relaxation to the endstate inside the light cone is $t^{-3}$ (this was proved in
\cite{christodoulou1} for the dispersive solutions and recently by Dafermos and Rodnianski
\cite{dr} for the collapsing solutions). In view of these  rigorous results, one might wonder
what is the point of studying this problem again. Our motivation is twofold.

First, it is natural to ask whether the decay rates mentioned above are optimal and, if so, what
are the corresponding amplitudes of the tails. This kind of \emph{quantitative} information might
be physically relevant (provided that the intuitions gained in this toy-model carry over to more
realistic situations), for example in assessing the possibility to detect the tails in future
gravitational wave experiments.

Second and foremost, we want to clarify some longstanding confusion which is widespread
throughout the vast relativity literature dealing with wave tails. To explain what this confusion
is about we need to make some historical remarks. The study of wave tails in general relativity
was launched in the seminal paper by Price \cite{price}, where he gave a heuristic argument that
a linear massless scalar field propagating on the fixed Schwarzschild background decays as
$t^{-3}$ near timelike infinity. This result has been later rederived by different methods and
confirmed numerically (the works \cite{leaver,gpp1,ching,barack,husa} are particularly
noteworthy), and finally proved rigorously in \cite{dr} (as a special decoupled case of the main
theorem on the coupled Einstein-scalar system).
An especially influential contribution to the study of tails was made in a pair of papers by
Gundlach, Price, and Pullin \cite{gpp1,gpp2}. In the first paper, \cite{gpp1}, they argued, using
linearized theory, that the $t^{-3}$ tail is due to the backscattering of the outgoing radiation
off the curvature at large distances, and therefore it is present for any asymptotically flat
solution, not only for black hole spacetimes. In the second paper, \cite{gpp2}, GPP solved the
spherically symmetric Einstein-scalar field equations numerically and found that, indeed, tails
do develop for any initial data  and moreover they decay as $t^{-3}$ regardless of the endstate
of evolution. This work was a significant step toward understanding of tails, however the fact
that it appeared back to back with \cite{gpp1} led also (somewhat ironically) to some confusion.
Namely, the remarkable agreement between the decay rates of tails observed numerically in the
nonlinear evolution \cite{gpp2} and the predictions of the linearized theory \cite{gpp1} has been
interpreted (first rather cautiously by the authors themselves and later with increasing sureness
in numerous citations of \cite{gpp2}) as if the linearized theory applies qualitatively (and, as
long as the power-law exponents of tails are concerned, even quantitatively) in the nonlinear
regime. We wish to point out that this interpretation is too naive.

We claim that the tails observed in \cite{gpp2} (and later confirmed in \cite{bo,cm} with better
numerical accuracy) have genuinely  \emph{nonlinear} origin for the dispersive solutions (while
for the collapsing solutions they have both linear and nonlinear ingredients).
 To substantiate
our claim we compute the late-time behavior (the decay rate \emph{and the amplitude} of a tail)
of the self-gravitating scalar field in even-dimensional spacetime  for small initial data using
the nonlinear perturbative scheme developed in our recent papers \cite{BCR1, BCR2, BCR3, BCR4,
nik}.
 The outcome of this simple analytic computation  is shown to agree extremely well with the results of
  high-precision numerical integration of the
 Einstein-scalar field equations in four and six dimensions, however it does \emph{not}
  agree with
 the linearized theory in dimensions higher than four. Thus, the equality of the decay rates of
linear and nonlinear tails seems to be a misleading idiosyncrasy of the Einstein-scalar field
equations in four dimensions.
  This paper is concerned
  only with subcritical initial data which lead to
 dispersion. Work on the collapsing solutions is still in progress.

The rest of the paper is organized as follows. In section~\ref{sec:eqs} we construct a simple
iterative scheme for solving the spherically symmetric Einstein-scalar field equations in
even-dimensional spacetimes. This scheme is applied in section~\ref{sec:pert3+1} to derive the
second-order approximation for the mass function and the third-order approximation for the scalar
field in four dimensions. The analogous calculation in higher even dimensions is done in
section~\ref{sec:pert5+1}. In Section~V we compare the nonlinear tails with the linear tails on
the fixed Schwarzschild background. Section~\ref{sec:numerics} contains numerical evidence
confirming the analytic formulae for the tail from sections~\ref{sec:pert3+1} and
\ref{sec:pert5+1}. In section~\ref{sec:conclusions} we make some final remarks.
\section{Field equations and the iterative scheme}
\label{sec:eqs} We consider the self-gravitating real massless scalar field $\phi$ in $d+1$
dimensions, where $d\geq 3$ is odd. The Einstein equations for the metric $g_{\alpha\beta} $ are
\begin{equation}\label{einstein}
    G_{\alpha\beta}=8\pi T_{\alpha\beta}\,,\qquad T_{\alpha\beta}=
    \nabla_{\alpha}\phi \nabla_{\beta}\phi-\frac{1}{2}g_{\alpha\beta}
    \left(\nabla_{\mu}\phi\nabla^{\mu}\phi\right)\,,
\end{equation}
and the scalar field satisfies the wave equation (which is equivalent to $\nabla_{\alpha}
T^{\alpha\beta}=0$)
\begin{equation}\label{we}
\nabla_{\mu} \nabla^{\mu}\phi=0\,.
\end{equation}
We assume spherical symmetry, so $\phi=\phi(t,r)$, and use the following ansatz for the metric
\begin{equation}
ds^2 = e^{2\alpha(t,r)}\left(-e^{2\beta(t,r)} dt^2 + dr^2\right) + r^2 d \Omega_{d-1}^2\,,
\end{equation}
where $d\Omega_{d-1}^2$ is the round metric on the unit  $(d-1)$--dimensional sphere. We define
also the mass function\\ $m(t,r)=(1-e^{-2\alpha}) r^{d-2}$. In these variables the Einstein
equations take the form
\begin{eqnarray}
\label{h-constraint} m'&=& \kappa\, r^{d-1} e^{-2\alpha} \left(\phi'^2 + e^{-2\beta}
\dot\phi^2\right),\hspace{1.2cm} \mbox{(Hamiltonian constraint)}
\\
\dot{m}&=& 2\kappa \,r^{d-1} e^{-2\alpha} \dot \phi \,\phi',\hspace{3.0cm} \mbox{(momentum
constraint)}
\\
\label{s-condition} \beta' &=& (d-2)\frac{m}{r^{d-1}} e^{2\alpha}\,,
\end{eqnarray}
where $\kappa=\dfrac{8 \pi}{d-1}$, and primes and dots denote partial derivatives with respect to
$r$ and $t$, respectively. Equation (6), corresponding to $G_t^t+G_r^r=8\pi(T_t^t+T_r^r)=0$, is
sometimes referred to as the polar slicing condition. The wave equation (\ref{we}), which can be
viewed as the integrability condition for equations (4) and (5), becomes
\begin{equation}\label{we2}
    \left(e^{-\beta}\dot\phi\right)^{\cdot}-\frac{1}{r^{d-1}}\left(r^{d-1} e^{\beta}
    \phi'\right)'=0\,.
\end{equation}
We assume that initial data are small, smooth, and compactly supported (the last assumption can
be replaced by a suitable fall-off condition)
\begin{equation}
\label{id} \phi(0,r) = \varepsilon f(r), \qquad \dot\phi(0,r) = \varepsilon g(r)\,.
\end{equation}
We make the following perturbative expansion
\begin{eqnarray}
m(t,r) &=& m_0(t,r) + \varepsilon m_1(t,r) + \varepsilon^2 m_2(t,r) + \dots,
\\
\beta(t,r) &=& \beta_0(t,r) + \varepsilon \beta_1(t,r) + \varepsilon^2 \beta_2(t,r) + \dots,
\\
\phi(t,r) &=& \phi_0(t,r) + \varepsilon \phi_1(t,r) + \varepsilon^2 \phi_2(t,r) + \varepsilon^3
\phi_3(t,r) + \dots.
\label{pertexpansion}
\end{eqnarray}
Substituting this expansion into the field equations and grouping terms with the same power of
$\varepsilon$ we get the iterative scheme which can be solved recursively.

 In this paper we consider perturbations about Minkowski spacetime, so $m_0=\beta_0=\phi_0=0$. At
the first order
 the metric functions $m_1=\beta_1=0$ (this follows from regularity at $r=0$), while $\phi_1$
 satisfies the flat space radial wave equation ($\Box=\partial_t^2-\partial_r^2-\dfrac{d-1}{r}\partial_r$)
\begin{equation}
\label{Box_phi1} \Box \phi_1 =0\,,\qquad \phi_1(0,r) = f(r)\,,\,\, \dot\phi_1(0,r) =  g(r)\,.
\end{equation}
The general spherically symmetric solution of equation (\ref{Box_phi1}) in odd spatial dimensions
$d=2\ell+3$ is given by a superposition of outgoing and ingoing waves
\begin{equation}
\label{phi1} \phi_1 (t,r) = \phi_1^{ret}(t,r) + \phi_1^{adv}(t,r)\,,
\end{equation}
where
\begin{equation}
\label{phi1retadv} \phi_1^{ret}(t,r) = \frac{1}{r^{\ell+1}}\,\sum_{k=0}^{l}  \frac {(2\ell-k)!}
{k!(\ell-k)!} \frac {a^{(k)}(u)}{(v-u)^{\ell-k}}\,, \qquad   \phi_1^{adv}(t,r) =
\frac{1}{r^{\ell+1}}\,\sum_{k=0} ^{\ell} (-1)^{k+1} \frac {(2\ell-k)!} {k!(\ell-k)!} \frac
{a^{(k)}(v)}{(v-u)^{\ell-k}}\,,
\end{equation}
and $u=t-r$, $v=t+r$ are the retarded and advanced times, respectively (the superscript in round
brackets denotes the $k$-th derivative). Note that for compactly supported initial data the
generating function $a(x)$ can be chosen to have compact support as well (this condition
determines $a(x)$ uniquely).

At the second order $\Box \phi_2 = 0$, hence $\phi_2=0$ (because it has zero initial data), while
the metric functions satisfy the following equations
\begin{eqnarray}
\label{m2prime3+1} m'_2 &=& \kappa\, r^{d-1} \left( \dot{\phi}_1^2 + \phi_1'^2 \right)\,,
\\
\label{m2dot3+1} \dot{m}_2 &=&  2 \kappa\, r^{d-1} \dot{\phi}_1 \phi'_1\,,
\\
\label{beta2prime3+1} \beta'_2 &=& \frac{(d-2) m_2}{r^{d-1}}\,.
\end{eqnarray}
We postpone the discussion of this system to the next section and proceed now to the third order,
where we have
\begin{equation}
\label{Box_phi3} \Box \phi_3  = 2 \beta_2 \ddot{\phi}_1 + \dot{\beta}_2 \dot{\phi}_1+ \beta'_2
\phi'_1.
\end{equation}
 To solve this equation we use the Duhamel formula for the solution of the
inhomogeneous wave equation $\Box \phi=N(t,r)$ with zero initial data
\begin{equation}
\label{duh} \phi(t,r)= \frac {1} {2 r^{\ell+1}} \int \limits_{0}^{t} d\tau \int
\limits_{|t-r-\tau|}^{t+r-\tau} \rho^{\ell+1} P_{\ell}(\mu)  N(\tau,\rho) d\rho\,,
\end{equation}
where $P_{\ell}(\mu)$ are Legendre polynomials of degree $\ell$ (recall that $\ell=(d-3)/2$) and
$\mu=(r^2+\rho^2-(t-\tau)^2)/2r\rho$ (note that $-1\leq \mu \leq 1$ within the integration
range). Applying this formula to equation (\ref{Box_phi3}), using null coordinates
$\eta=\tau-\rho$ and $\xi=\tau+\rho$, and the abbreviation $K(\beta,\phi)=2 \beta \ddot{\phi} +
\dot{\beta} \dot{\phi} + \beta' \phi'$, we obtain
\begin{equation}
\label{phi3} \phi_3(t,r) = \frac {1} {2^{\ell+3} r^{\ell+1}} \int\limits_{|t-r|}^{t+r} d\xi \int
\limits_{-\xi}^{t-r} (\xi-\eta)^{\ell+1} P_{\ell}(\mu) K(\beta_2(\xi,\eta),\phi_1(\xi,\eta))
d\eta\,,
\end{equation}
where now $ \mu=(r^2+(\xi-t)(t-\eta))/r(\xi-\eta)$. If the initial data (\ref{id}) vanish outside
a ball of radius $R$, then for $t>r+R$ we may drop the advanced part of $\phi_1(t,r)$ and
interchange the order of integration in (\ref{phi3}) to get
\begin{equation}
\label{phi3(2)} \phi_3(t,r) = \frac {1} {2^{\ell+3} r^{\ell+1}} \int \limits_{-\infty}^{\infty}
d\eta \int \limits_{t-r}^{t+r} (\xi-\eta)^{\ell+1} P_{\ell}(\mu)
K(\beta_2(\xi,\eta),\phi^{ret}_1(\xi,\eta)) \, d\xi \,.
\end{equation}
In order to determine the behavior of $\phi_3(t,r)$ for large $t$ we need only to know the
behavior of the metric function $\beta_2(t,r)$ along the light cone for large values of $r$ (the
intersection of the integration range in (\ref{phi3(2)}) with the support of
$\phi_1^{ret}(t,r)$). This calculation will be done in the next section.  Having that, we shall
expand the function $K$ in (\ref{phi3(2)}) in the inverse powers of $(\xi-\eta)$ and use the
identity (see the appendix in \cite{BCR4} for the derivation)
\begin{equation}\label{master}
\int \limits_{t-r}^{t+r} d\xi \, \frac {P_{\ell} (\mu)} {(\xi-\eta)^{n}} =  (-1)^{\ell} \frac
{2(n-2)^{\underline{\ell}}} {(2\ell+1)!!}  \frac {r^{\ell+1} (t-\eta)^{n-\ell-2}} {[(t-\eta)^2 -
r^2]^{n-1}} \, F \left( \left. \begin{array}{c} \frac {\ell+2-n} {2}, \, \frac {\ell+3-n} {2}  \\
\ell + 3/2
\end{array} \right| \left( \frac {r} {t - \eta} \right)^{2} \right)\,,
\end{equation}
where $(n-2)^{\underline \ell}=(n-2)(n-3)\cdots(n-\ell-1)$ ($\ell>0$) and $(n-2)^{\underline
0}=1$.

 If one has no fear, this iterative procedure can  be continued to higher orders,
  however it seems like an overkill in
 view of two facts. First, the iteration has no chance to converge (\emph{cf.}\cite{alan}) so our
 perturbation series is only asymptotic. Second, already the third-order approximation shows
 excellent
 agreement with numerical results
 (see section~VI).
\section{Nonlinear tail in $3+1$ dimensions}
\label{sec:pert3+1} In this section, written mainly for the benefit of the reader who is not
interested in higher dimensions,  we follow the general strategy sketched above to present a
detailed calculation of the third-order iterate $\phi_3(t,r)$ in three spatial dimensions. In the
next section we shall repeat this calculation for any odd spatial dimension $d\geq 3$.

For $d=3$ (hence $l=0$), the solution (\ref{phi1},\ref{phi1retadv}) of the free wave equation
takes the form
\begin{equation}\label{phil=0}
    \phi_1(t,r)=\frac{a(u)-a(v)}{r}\,,
\end{equation}
where the function $a(x)$ is uniquely determined by the initial data. Substituting (\ref{phil=0})
into (\ref{m2prime3+1}) and integrating, we get
\begin{equation}\label{m2f}
m_2(t,r) \stackrel{t>R}{=} 4\pi \int \limits_0^r \left( 2 a'^2(t-\rho) - \partial_{\rho} \frac
{a^2(t-\rho)} {\rho} \right)\, d\rho \,,
\end{equation}
where we used that $m_2(t,r=0)=0$, which follows from regularity of initial data at the origin
and (\ref{m2dot3+1}). Here and in the following we use repeatedly the fact that $a(x)=0$ for
$|x|>R$, $R$ being the radius of a ball on which the initial data (\ref{id}) are supported. To
describe the behavior of $m_2(t,r)$ near the lightcone it is convenient to use the null
coordinate $u=t-r$ instead of $t$, and rewrite (\ref{m2f}) as
\begin{equation}
m_2(u,r) \stackrel{r+u>R}{=} 4\pi \left( 2 \int \limits_u^\infty a'^2(s) \, ds - \frac {a^2(u)}
{r}\right)\,.
\end{equation}
 Next, using the gauge freedom to set $\beta_2(t,r=0)=0$ and integrating equation
 (\ref{beta2prime3+1}), we get
\begin{equation}
\beta_2(t,r) \stackrel{t>R}{=} 4\pi \left( 2 \int \limits_0^r \frac {1} {\rho^2} \int
\limits_{t-\rho}^{\infty} a'^2(s) \, ds \, d\rho - \int \limits_0^r \frac {a^2(t-\rho)} {\rho^3}
\, d\rho \right) \, .
\end{equation}
The first integral can be integrated by parts giving
\begin{equation}
\beta_2(u,r) \stackrel{r+u>R}{=} 4\pi \left( -\frac{2}{r} \int \limits_u^{\infty} a'^2(s) \, ds +
2 \int \limits_{u}^{\infty} \frac {a'^2(s)} {r-(s-u)} \, ds - \int \limits_{u}^{\infty} \frac
{a^2(s)} {(r-(s-u))^3} \, ds \right) \, .
\end{equation}
In order to determine the tail of $\phi_3$ we need only two leading terms in the expansion of the
above formula in the inverse powers of $r$. Hereafter, it is convenient to define the following integrals
(for non-negative integers $m,n$)
\begin{equation}
\label{Imn} I^{m}_{n} (u) = \int \limits_u^\infty (s-u)^m \left(a^{(n)}(s)\right)^2 \, ds.
\end{equation}
Then our results can be cast in the form:
\begin{eqnarray}
\label{beta2} \beta_2(u,r) &\stackrel{r+u>R}{=}& \frac {4\pi} {r^2} \left[ 2 I^{1}_{1}(u) + \frac{1}{r}
(2 I^{2}_{1}(u) - I^{0}_{0}(u)) + \mathcal{O} \left( \frac {1} {r^2} \right) \right]\,,
\\
\label{beta2dot} \dot{\beta}_2(u,r) &\stackrel{r+u>R}{=}& - \frac {4\pi} {r^2} \left[ 2 I^{0}_{1}(u) +
\frac{1}{r} (4 I^{1}_{1}(u) - a^2(u)) + \mathcal{O} \left( \frac {1} {r^2} \right) \right]\,,
\\
\label{beta2prime} \beta_2'(u,r)  &\stackrel{r+u>R}{=}& \frac {4\pi} {r^2} \left[ 2 I^{0}_{1}(u) -
\frac{1}{r} a^2(u) + \mathcal{O} \left( \frac {1} {r^2} \right) \right]\,.
\end{eqnarray}
Substituting (\ref{phil=0}) and (\ref{beta2}-\ref{beta2prime}) into (\ref{phi3(2)}) (with
$\ell=0$) we get for $t>r+3R$
\begin{equation}
\label{phi3(3)} \phi_3(t,r) = \frac {2^4 \pi} {r} \int \limits_{-\infty}^{+\infty} d\eta \int
\limits_{t-r}^{t+r} \frac {d\xi} {(\xi-\eta)^2}  \left[ \frac {d} {d \eta} \left( I^{1}_{1}(\eta)
a'(\eta) \right) + \frac{1}{\xi-\eta} \left(I^{0}_{1}(\eta) a(\eta) \!+\! \frac {d} {d \eta}
Q_0(\eta) \right)\, \! + \!\mathcal{O} \left( \frac {1} {(\xi-\eta)^2} \right) \right],
\end{equation}
where
\begin{equation}\label{Q_0}
Q_0(\eta) = 2 I^{1}_{1}(\eta) a(\eta) + (2 I^{2}_{1}(\eta) - I^{0}_{0}(\eta)) a'(\eta)\,.
\end{equation}
Elementary integrations over $\xi$ and  by parts over $\eta$ yield for large retarded times
$u=t-r$
\begin{equation}
\label{phi3tail} \phi_3(t,r) = \frac {t}{(t^2-r^2)^2}\left[\Gamma_0 + \mathcal{O} \left( \frac
{t} {t^2-r^2} \right)\right]\, ,
\end{equation}
where the coefficient
\begin{equation}
\Gamma_0 = -2^5\pi \int \limits_{-\infty}^{+\infty} I^{0}_{1}(s) a(s)\, ds
\end{equation}
is the only trace of initial data. From (\ref{phi3tail}) we obtain the late-time tails in two
asymptotic regimes: $\phi_3(t,r)\simeq \Gamma_0 t^{-3}$ at future timelike infinity
($r=const,t\rightarrow\infty$) and $(r\phi_3)(v=\infty,u)\simeq \dfrac{1}{4}\Gamma_0 u^{-2}$
along future null infinity ($v=\infty, u\rightarrow \infty$).

\section{Nonlinear tail in $d+1$ dimensions}
\label{sec:pert5+1} Proceeding along the same lines as described in detail in the previous
section, we get the analogues of formulae (\ref{beta2}-\ref{beta2prime}) in $d+1$ dimensions
(recall that $d=2\ell+3$ so $\kappa=4\pi/(\ell+1)$)
\begin{eqnarray}
\label{beta2(d+1)} \beta_2(u,r) &\stackrel{r+u>R}{=}& \frac {(2\ell+1) \kappa} {r^{2\ell+2}}
\left[ 2 I^{1}_{\ell+1}(u) + \frac{\ell+1}{r} (2 I^{2}_{\ell+1}(u) - (\ell+1) I^{0}_{\ell}(u)) + \mathcal{O} \left(
\frac {1} {r^2} \right) \right]\,,
\\
\label{beta2dot(d+1)} \dot{\beta}_2(u,r) &\stackrel{r+u>R}{=}& - \frac {(2\ell+1) \kappa}
{r^{2\ell+2}} \left[ 2 I^{0}_{\ell+1}(u) + \frac{\ell+1}{r} (4 I^{1}_{\ell+1}(u) - (\ell+1)
\left(a^{(\ell)}(u)\right)^2) + \mathcal{O} \left( \frac {1} {r^2} \right) \right],
\\
\label{beta2prime(d+1)} \beta_2'(u,r)
%= \frac {(d-2) m_2(u,r)} {r^{d-1}}
&\stackrel{r+u>R}{=}& \frac {(2\ell+1) \kappa} {r^{2\ell+2}} \left[ 2 I^{0}_{\ell+1}(u) - (\ell+1)^2
\frac{1}{r} \left(a^{(\ell)}(u)\right)^2 + \mathcal{O} \left( \frac {1} {r^2} \right) \right]\,,
\end{eqnarray}
with $I^{m}_{n}(u)$ defined in (\ref{Imn}). Substituting these expressions into (\ref{phi3(2)})
we get for $t>r+3R$
\begin{eqnarray}
\label{phi3(3)_l}
\phi_3(t,r) &=& \frac {2^{2\ell+2} (2\ell+1) \kappa} {r^{\ell+1}} \int \limits_{-\infty}^{+\infty} d\eta \int
\limits_{t-r}^{t+r} \frac {d\xi} {(\xi-\eta)^{2 \ell + 2}}  \left[ \frac {d} {d \eta} \left( I^{1}_{\ell+1}(\eta)
a^{(\ell+1)}(\eta) \right) \right.
\nonumber\\
&& + \left. \frac{\ell+1}{\xi-\eta} \left(I^{0}_{\ell+1}(\eta) a^{(\ell)}(\eta) \!+\! \frac {d} {d \eta} Q_{\ell}(\eta)
\right)\, \! + \!\mathcal{O} \left( \frac
{1} {(\xi-\eta)^2} \right) \right],
\end{eqnarray}
where
\begin{equation}
\label{Q_l}
Q_{\ell}(\eta) = (\ell+2) I^{1}_{\ell+1}(\eta) a^{(\ell)}(\eta) + (2 I^{2}_{\ell+1}(\eta) - (\ell+1) I^{0}_{0}(\eta)) a^{(\ell+1)}(\eta)\,.
\end{equation}
Using the identity (\ref{master}) we get the asymptotic behavior  for large retarded times
$u=t-r$
\begin{equation}
\label{phi3tail(d+1)} \phi_3(t,r)\! =\! \frac {t^{\ell+1}} {(t^2-r^2)^{2\ell+2}} \left\{ \left(
4\ell+2- \ell (1-\frac{r^2}{t^2}) \right)
F\! \left(\! \!\left. \begin{array}{c} - \frac {\ell} {2}, \, - \frac {\ell} {2} + \frac {1} {2} \! \\
\ell + 3/2 \end{array} \right| \frac {r^2} {t^2}\! \right)
- \!(2\ell+1) F\! \left(\!\! \left. \begin{array}{c} - \frac {\ell} {2}, \, - \frac {\ell} {2} - \frac {1} {2}  \\
\ell + 3/2 \end{array} \right| \frac{r^2} {t^2}\!  \right) \right\}
\left[\frac{\Gamma_{\ell}}{\ell+1}\! + \!\mathcal{O} \left( \!\frac {t} {t^2-r^2}\!
\right)\right]\, ,
\end{equation}
where
\begin{equation}\label{gamma_l}
\Gamma_\ell = (-1)^{l+1} 2^{3l+5} \pi \int \limits_{-\infty}^{+\infty} I^{0}_{\ell+1}(s)
a^{(\ell)}(s)\, ds\,.
\end{equation}
For $\ell=0$ the formula  (\ref{phi3tail(d+1)}) reduces to (\ref{phi3tail}). For $\ell\geq1$ the
integral (\ref{gamma_l}) can be integrated by parts again, giving
\begin{equation}
\Gamma_\ell = (-1)^{l+1} 2^{3l+5} \pi \int \limits_{-\infty}^{+\infty}
\left(a^{(\ell+1)}(s)\right)^2 a^{(\ell-1)}(s)\, ds.
\end{equation}
Asymptotics at time and  null infinity are easily obtained from (\ref{phi3tail(d+1)}). They read:
\begin{eqnarray}
\phi_3(t,r) &=& \frac {1} {t^{3\ell+3}} \left[\Gamma_{\ell}
 + \mathcal{O} \left( \frac {1} {t} \right) \right] \,,\\
  (r^{\ell+1}\phi_3)(v=\infty,u) &=& \frac{1}{u^{2\ell+2}}
  \left[\frac{(2\ell+1)!(2\ell+1)!!}{2(3\ell+2)!} \Gamma_{\ell} + \mathcal{O} \left( \frac {1} {u}
  \right) \right]\,.
\end{eqnarray}
This is our main result. We claim that the formulae (44) and (45) provide very good
approximations of the tail for solutions with sufficiently small initial data. By this we mean
that for any given smooth compactly supported profiles $f(r)$ and $g(r)$ in (8) one can choose
$\varepsilon$ so small that
\begin{eqnarray}
\lim_{t\rightarrow\infty} t^{3\ell+3}
|\phi(t,r)-\varepsilon^3\phi_3(t,r)|&=&\mathcal{O}(\varepsilon^5)\,,\\
\lim_{u\rightarrow\infty} u^{2\ell+2}
|(r^{\ell+1}(\phi-\varepsilon^3\phi_3))(v=\infty,u))|&=&\mathcal{O}(\varepsilon^5)\,,
\end{eqnarray}
at time and  null infinity, respectively. Numerical evidence for this claim is given in
Section~VI.
\section{Linear scalar waves on Schwarzschild background}
For the sake of completeness, in this section we recall briefly what is known about the decay of
the massless scalar field propagating outside the $d+1$ dimensional Schwarzschild black hole
\begin{equation}\label{sch}
    ds^2=-\left(1-\frac{M}{r^{d-2}}\right) dt^2 +
    \left(1-\frac{M}{r^{d-2}}\right)^{-1}
     dr^2 + r^2
    d\Omega_{d-1}^2\,.
\end{equation}
As above, we consider only odd spatial dimensions $d\geq 3$ and use the integer index
$\ell=(d-3)/2$.
 In terms
of the tortoise coordinate $x$, defined by $dr/dx=1-M/r^{2\ell+1}$, and the variable
$\psi(x)=r^{\ell+1}\phi(r)$, the radial wave equation in the metric (\ref{sch}) for $r\geq M$
reduces to the flat spacetime $1+1$ dimensional wave equation with the potential (on the whole
axis $-\infty<x<\infty$)
\begin{equation}\label{eqsch}
    \partial_t^2\psi -\partial_x^2 \psi +V(x) \psi=0, \qquad
    V=\left(1-\frac{M}{r^{2\ell+1}}\right) \left(\frac{\ell(\ell+1)}{r^2}+\frac{M (\ell+1)^2}{
    r^{2\ell+3}}\right)\,.
\end{equation}
Now, there is an important difference between $\ell=0$ and $\ell>0$ cases which is due to the
fact that only for $\ell=0$ the tortoise coordinate involves the logarithm. More concretely, for
$\ell=0$ we have
\begin{equation}\label{x3}
    x=r+M \ln(r/M-1)\,,
\end{equation}
hence for $x\gg M$
\begin{equation}\label{x3exp}
    r=x-M\ln(x/M)+\frac{M^2 \ln(x/M)}{x}+\frac{M^2}{x}+\mathcal{O}
    \left(\frac{M^3\ln^2(x/M)}{x^2}\right)\,,
\end{equation}
and therefore
\begin{equation}\label{Vx3}
    V(x) = \frac{M}{x^3}+\frac{3M^2\ln(x/M)}{x^4}-\frac{M^2}{x^4}+\mathcal{O}
    \left(\frac{M^3\ln^2(x/M)}{x^5}\right)\,,
\end{equation}
which gives rise to the Price tail $\phi(t,r)\sim M t^{-3}$ \cite{price}.

 In contrast, for $\ell\geq 1$ we have (see
\cite{BCR3})
\begin{equation}\label{exp2}
    r =
    x+\frac{1}{2\ell}\frac{M}{x^{2\ell}}-\frac{2\ell+1}{2\ell(4\ell+1)}\frac{M^2}{x^{4\ell+1}}
    +\mathcal{O} \left( \frac{M^3}{x^{6\ell+2}}\right)\,,
\end{equation}
which implies that for large $x$
\begin{equation}\label{Vxd}
    V(x) = \frac{\ell(\ell+1)}{x^2}+\frac{(2\ell+1)^2(\ell+1)(4\ell+3)}{4 \ell(4\ell+1)} \frac{M^2}{x^{4\ell+4}}
    +\mathcal{O} \left( \frac{M^3}{x^{6\ell+5}}\right)\,.
\end{equation}
A remarkable feature of this effective potential  is the absence of a term proportional to~$M$.
It is for this reason that the tail drops very rapidly:
\begin{equation}\label{tailSd}
    \phi(t,r)\sim \frac{M^2}{t^{6\ell+4}} \,,
\end{equation}
as follows from the general formula $t^{-(\alpha+2\ell)}$ for the tail produced by the potential
of the form $\ell(\ell+1)/x^2+U(x)$ with $U(x)\sim x^{-\alpha}$ for large $x$ \cite{ching,BCR3}.

\addtolength{\topmargin}{-1.0pc} \addtolength{\textheight}{2.0pc}
\section{Numerics}
\label{sec:numerics} In order to verify the above analytic predictions we solved numerically the
initial value problem (\ref{h-constraint}-\ref{id}) for various initial data. To this end we
rewrite the wave equation (\ref{we2}) as a pair of two first order equations for auxiliary fields
$\Phi = \phi'$ and $\Pi=e^{-\beta}\dot{\phi}$:
\begin{equation}\label{we-num}
\dot\Pi = \frac{1}{r^{d-1}}\left(r^{d-1} e^{\beta} \Phi \right)', \quad \mbox{and} \quad \dot\Phi = \left(e^{\beta} \Pi \right)'.
\end{equation}
We solve these equations with fourth-order accurate Runge-Kutta time stepping using finite
differencing in space.  At each time step we update the metric functions $m(t,r)$ and
$\beta(t,r)$ by integrating the hamiltonian constraint
  (\ref{h-constraint}) and the slicing condition (\ref{s-condition}) with fourth-order Runge-Kutta method.
   To ensure regularity at the origin we impose the boundary conditions
$\Phi(t,0)=0$ and $\Pi'(t,0)=0$, which are implemented by $\Phi(t,r)$ and $\Pi(t,r)$ being odd
and even functions of $r$, respectively.
To avoid the contamination of the tail by spurious reflections from the outer boundary of the
computational grid we place that boundary far away and compute the solution only inside the
domain of dependence of the initial surface. As was pointed out in \cite{ching}, a reliable
numerical computation of tails
  requires high-order finite difference schemes, since otherwise the ghost potentials generated by
  discretization errors produce artificial tails which might mask the genuine behavior.
  We used fourth and tenth-order difference schemes for $d=3$ and $d=5$ dimensions,
  respectively\footnote{On the fixed Schwarzschild background in $d=5$ the scalar field
  $\phi(t,r)$ decays as
  $t^{-10}$ (see (\ref{tailSd})). If a weak self-gravitating scalar field decayed at this rate,
  its tail would be hidden under an artificial tail generated by a ghost potential unless
  the tenth or higher order discretization is used.}.
  To eliminate high-frequency numerical instabilities we add a small amount of
  artificial dissipation \cite{kreiss-oliger}, that is after each time step advancing solution $f$
  from $t$ to $t+\Delta t$ on a grid with $(\Delta t,\Delta r$) mesh
  we add the Kreiss-Oliger dissipative term
$f(t+\Delta t,r) \longrightarrow f(t+\Delta t,r) + Q_k\,f(t,r)$, where (for consistency with
$2(k-1)$-order finite difference scheme) $Q_k$ is a finite-difference operator of order $2k$ of
the form $Q_k = (-1)^{k+1} \frac {\sigma} {2^{2k}} \left( \frac {\Delta t} {\Delta r} \right)\,
\left(\Delta_{+}\right)^k\, \left(\Delta_{-}\right)^k$ where $\sigma$ is of order of $1$ and
$\Delta_{\pm} f(t,r) = \pm (f(t,r \pm \Delta r) - f(t,r))$. Finally, to suppress the accumulation
of round-off errors at late times our codes were run in 128-bit precision. For the above reasons
the accurate numerical simulations of tails, albeit straightforward, are computationally
expensive even in spherical symmetry.

The numerical results presented here correspond to initial data generated by the function (see
(\ref{phi1retadv}))
\begin{equation}\label{idn}
\varepsilon a(x) = \frac {\varepsilon} {\sqrt{2\pi}} \exp\left(-x^2\right)
\end{equation}
for different values of $\varepsilon$. For these initial data our third-order approximation (44)
yields the following asymptotic behavior at timelike infinity
\begin{equation}
\dot\phi(t,r) = 12 \sqrt{\pi} \, \varepsilon^3 \, \frac{1}{t^4} \left(1 + \mathcal{O}
\left(\frac{1}{t} \right) \right)\qquad \mbox{for}\quad d=3\,,
\end{equation}
and
\begin{equation}
\dot\phi(t,r) = - \frac {1024 \sqrt{2} \, \varepsilon^3} {\sqrt{3}} \, \frac{1}{t^7} \left(1 +
\mathcal{O} \left(\frac{1}{t} \right) \right)\qquad \mbox{for}\quad d=5\,.
\end{equation}
In Fig. 1 we plot $\dot \phi(t,0) = e^{\beta} \Pi(t,0)$ in $d=3$ and $d=5$ for three different
values of $\varepsilon$. The late time tails are clearly seen as straight lines on log-log plots.
We fit our numerical data with the formula
\begin{equation}
\label{numphidot} \dot \phi (t,r) = A t^{-\gamma} \exp \left(B/t + C/t^2 \right)\,,
\end{equation}
which gives the local power index (LPI) \cite{bo}
\begin{equation}
\label{lpi} n(t,r) := -t \ddot \phi(t,r) / \dot\phi(t,r) = \gamma + B/t + 2C/t^2\,.
\end{equation}
 We plot the local power index at $r=0$ as a function of $1/t$ in Fig.~2.
\begin{figure}[h]
\begin{tabular}{cc}
\includegraphics[width=0.45\textwidth]{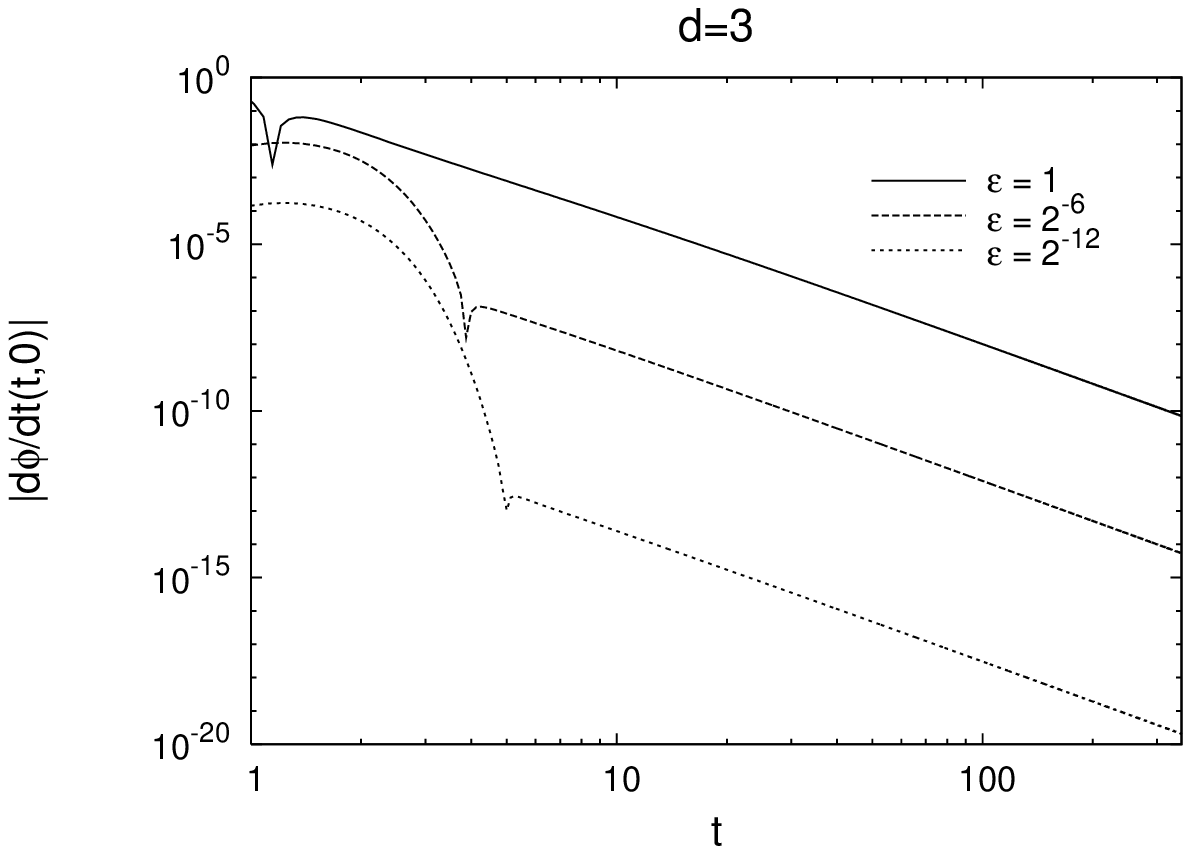}
&
\includegraphics[width=0.45\textwidth]{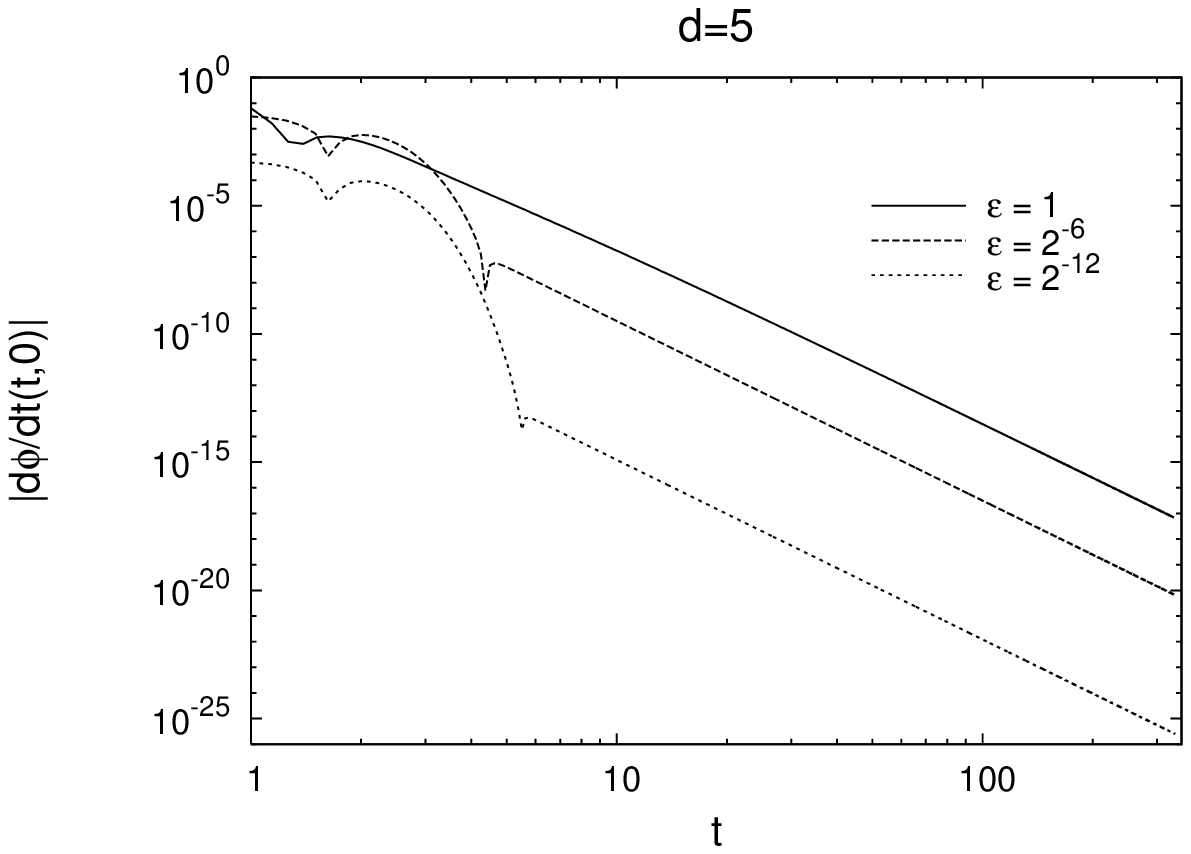}
\\
\end{tabular}
\caption{\small{The log-log plot of $\dot\phi(t,0)$ for small ($\varepsilon=2^{-12}$),
intermediate ($\varepsilon=2^{-6}$), and large ($\varepsilon=1$) amplitudes of initial data. The
slopes ($\gamma=4$ for $d=3$ and $\gamma=7$ for $d=5$) do not depend on the size of the data. }}
\label{fig.tails}
\end{figure}
\begin{figure} [t]
\begin{tabular}{cc}
\includegraphics[width=0.45\textwidth]{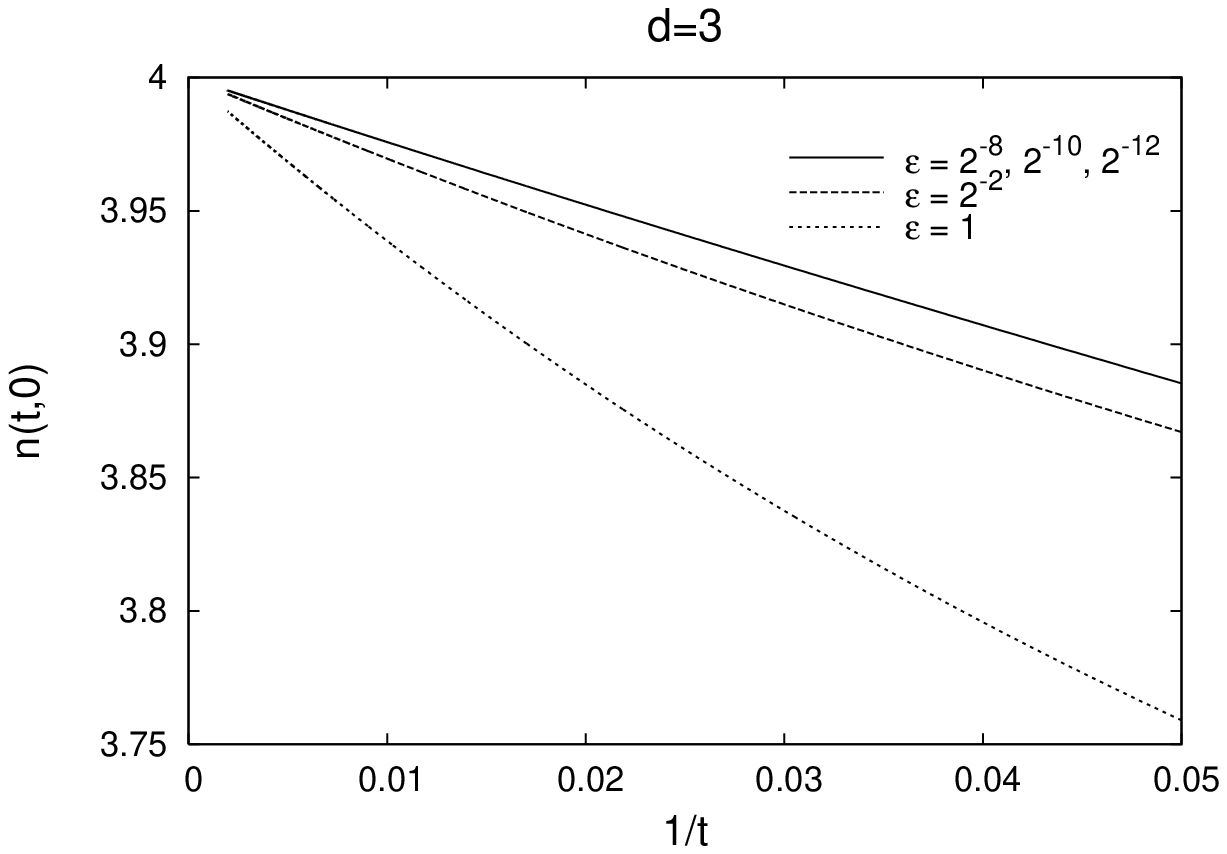}
&
\includegraphics[width=0.45\textwidth]{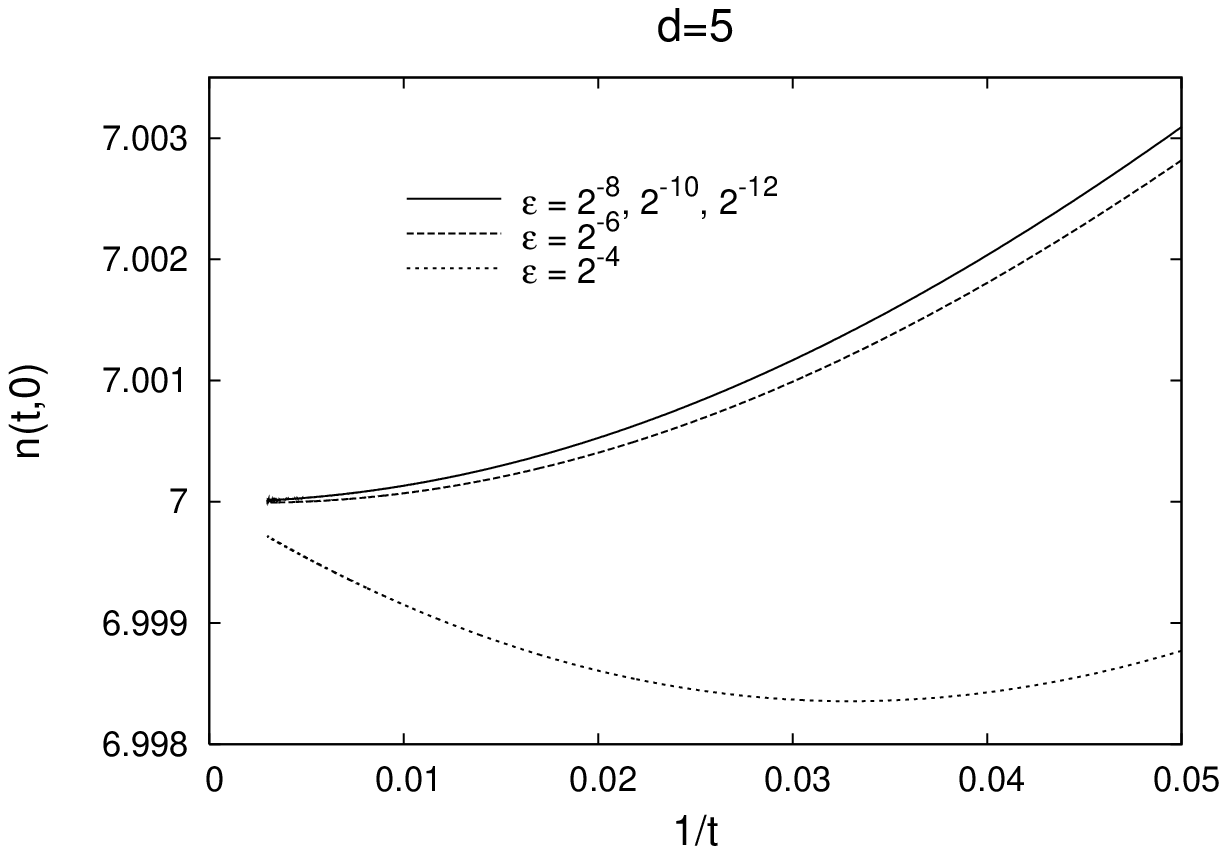}
\\
\end{tabular}
\caption{\small{The local power index $n(t,0)$ as a function of $1/t$. The curves corresponding
to small initial data ($\varepsilon=2^{-8},2^{-10},2^{-12}$) are indistinguishable which
indicates that higher order corrections in the perturbation series are negligible. }}
\label{fig.local}
\end{figure}
  Our fitting procedure proceeds in two steps. First,
from the local power index data on the interval $0 < 1/t <1/50$ we fit $\gamma$, $B$ and $C$ in
(\ref{lpi}). Next, having determined $\gamma$, $B$ and $C$ in this way, we fit $A$ in
(\ref{numphidot}) from $\dot \phi$ data on the interval $t>50 $. We have verified that the
outcome of the fit (the amplitude $A$ and the decay rate $\gamma$) does not depend on the
observation point $r$.
 The results for $r=0$ and their confrontation with analytic predictions are
summarized in Table~1 for $d=3$ and Table~2 for $d=5$.  The agreement between our third-order
approximation and the results of numerical integration of the Einstein-scalar field equations
 is excellent for sufficiently small initial data.

In Fig.~3 we plot the fitted amplitude of the tail versus the amplitude of initial data and
compare it with
 our third-order analytic formula. The deviation from the scaling $A\sim \varepsilon^3$ for large
 $\varepsilon$ signals the breakdown of the third-order approximation. We stress that we get the same
  decay rates ($\gamma=4$ for $d=3$ and $\gamma=7$ for $d=5$) for
all subcritical evolutions, regardless of whether our third-order formula predicts correctly the
amplitude of the tail (for small data) or fails (for large data
 where higher-order terms in the asymptotic expansion (\ref{pertexpansion}) cannot be neglected).
\begin{figure}[h]
\begin{tabular}{cc}
\includegraphics[width=0.45\textwidth]{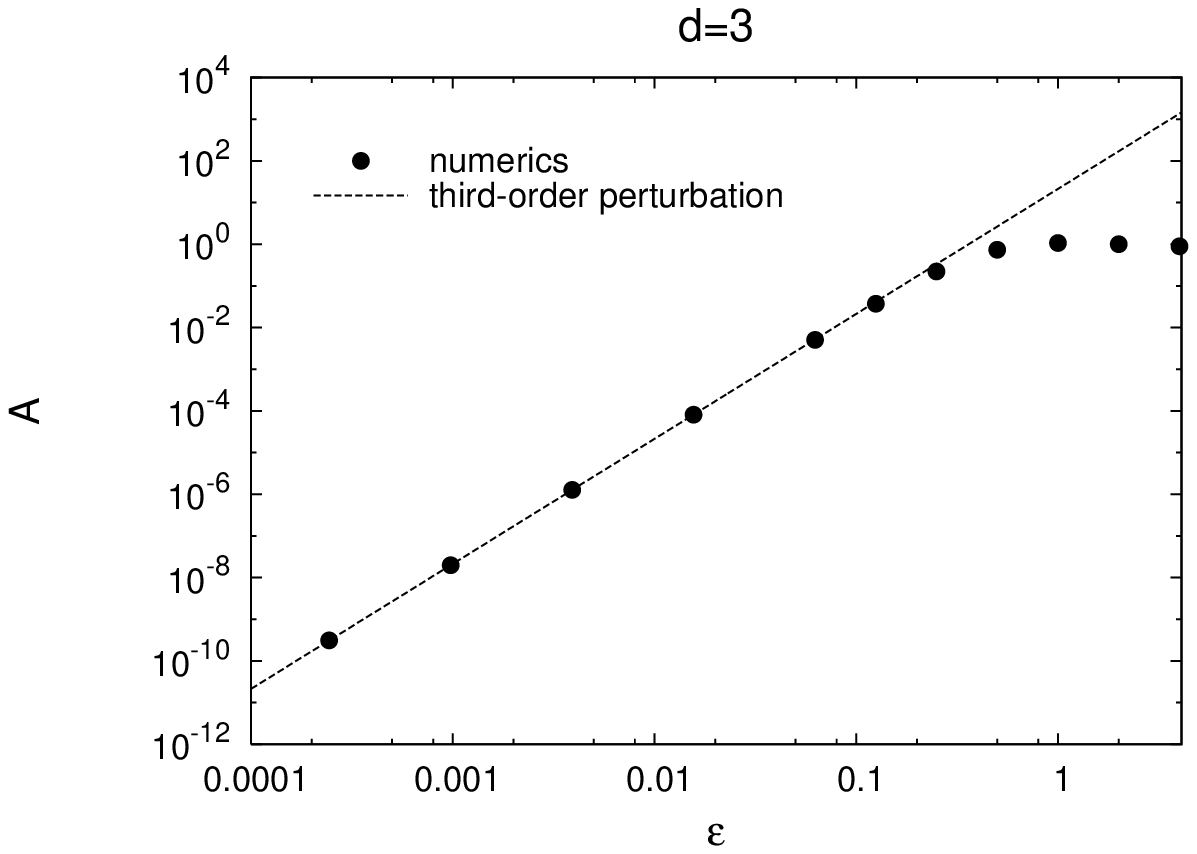}
&
\includegraphics[width=0.45\textwidth]{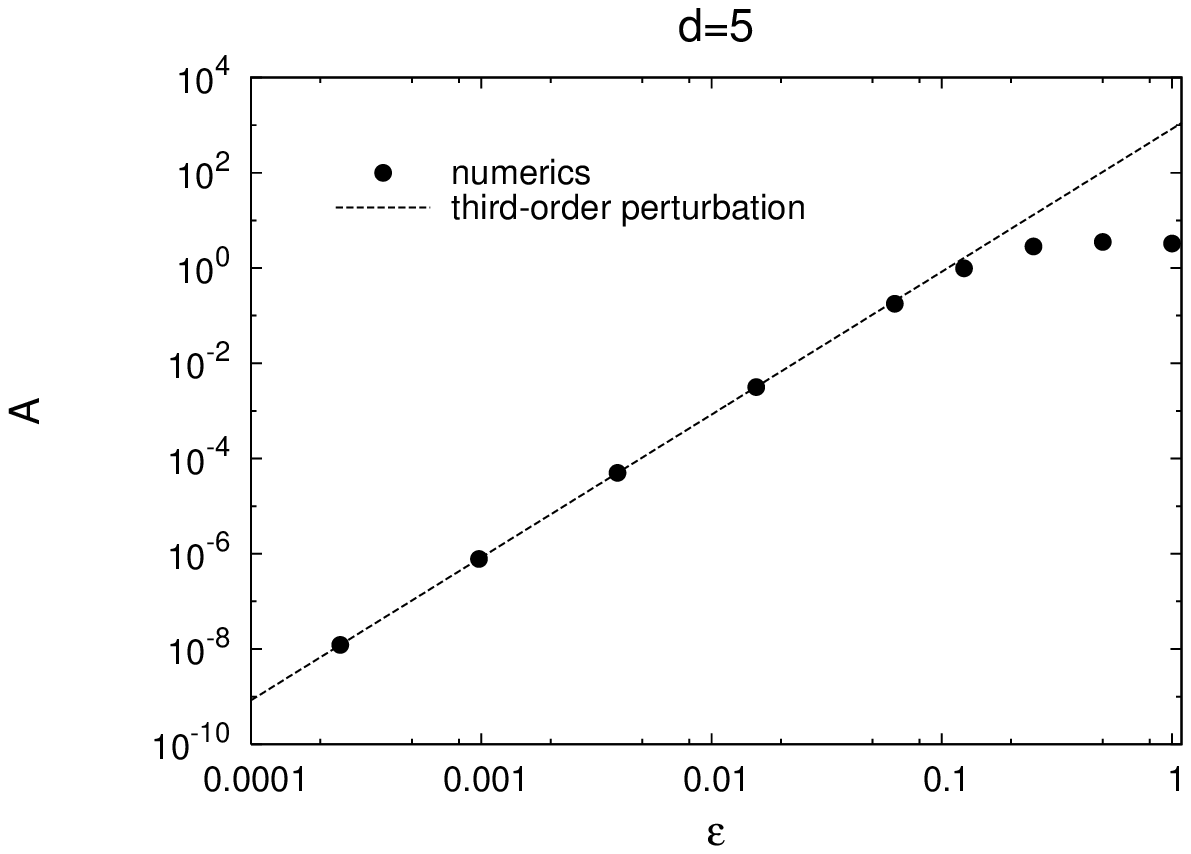}
\\
\end{tabular}
\caption{\small{The log-log plot of the amplitude of the tail as a function of the amplitude of
initial data (black dots). The third-order approximation (dashed line) is excellent for small
data, but it deteriorates for large data lying near the threshold for black hole formation
($\varepsilon\sim 1$). The scaling $A\sim \varepsilon^3$ was previously observed by GPP
 (see Fig.~14 in \cite{gpp2}).}} \label{fig.ampl}
\end{figure}

\begin{table}[ht]
\centering \caption{$d=3$.} \label{tableD3} \vspace{0.2cm}
\begin{tabular}{|c||c|c|c||c|c||c|}
\hline $\varepsilon$ & \multicolumn{3}{c||}{Numerics: LPI data} & \multicolumn{2}{c||}{Theory
(third order)}
& Numerics: $\dot \phi$ data \\
\cline{2-7}
 & $B$ & $C$ & $\gamma$ & $\gamma$  & $A$  & $A$ \\
\hline \hline
$\,\,2^{-12}\,\,$ & $\,\,$-2.45384$\,\,$ & $\,\,1.98180\,\,$ & $\,\,4.0000\,\,$ & $\,\,\,\,4\,\,\,\,$ & 3.09511e-10 & 3.09511e-10 \\
\hline
$2^{-10}$ & -2.44983 & 1.71398 & 4.0000 & 4 & 1.98087e-08 & 1.98083e-08 \\
\hline
$2^{-8}$ & -2.44977 & 1.69814 & 4.0000 & 4 & 1.26776e-06 & 1.26760e-06 \\
\hline
$2^{-6}$ & -2.45270  & 1.71275 & 4.0000 & 4 & 8.11365e-05 & 8.09971e-05 \\
\hline
$2^{-4}$ & -2.49938 & 1.95863 & 4.0000 & 4 & 5.19274e-03 & 5.05355e-03 \\
\hline
$2^{-3}$ & -2.64286 & 2.70811 & 4.0000 & 4 & 0.0415419 & 0.0373293 \\
\hline
$2^{-2}$ & -3.14114 & 5.24492 & 4.0000 & 4 & 0.332335 & 0.222460 \\
\hline
$2^{-1}$ & -4.42597 & 11.2084 & 4.0000 & 4 & 2.65868 & 0.737111 \\
\hline
$1$      & -6.49635 & 18.7893 & 3.9999 & 4 & 21.2694 & 1.07316 \\
\hline
$2$      & -8.98950 & 25.4951 & 4.0002 & 4 & 170.156 & 0.997247 \\
\hline
$4$      & -11.8828 & 32.3576 & 4.0021 & 4 & 1361.24 & 0.892345 \\
\hline
\end{tabular}
\end{table}

\begin{table}[ht]
\centering \caption{$d=5$.} \label{tableD6} \vspace{0.2cm}
\begin{tabular}{|c||c|c|c||c|c||c|}
\hline $\varepsilon$ & \multicolumn{3}{c||}{Numerics: LPI data} & \multicolumn{2}{c||}{Theory
(third order)}
 & Numerics: $\dot \phi$ data \\
\cline{2-7}
 & $B$ & $C$ & $\gamma$  & $\gamma$  & $A$ & $A$ \\
\hline \hline
$\,\,2^{-12}\,\,$ & $\,\,$4.64290e-04$\,\,$ & $\,\,0.650257\,\,$ & $\,\,7.0000\,\,$ & $\,\,\,7\,\,\,$ & -1.21667e-08 & -1.21667e-08 \\
\hline
$2^{-10}$ & 2.48554e-04 & 0.654252 & 7.0000 & 7 & -7.78672e-07 & -7.78644e-07 \\
\hline
$2^{-8}$ & -1.02440e-04 & 0.654138 & 7.0000 & 7 & -4.98350e-05 & -4.98072e-05 \\
\hline
$2^{-6}$ & -6.10626e-03 & 0.660806 & 7.0000 & 7 & -3.18944e-03 & -3.16123e-03 \\
\hline
$2^{-4}$ & -0.100262 & 0.765030 & 7.0000 & 7 & -0.204124 & -0.177845 \\
\hline
$2^{-3}$ & -0.380511 & 1.07174 & 7.0000 & 7 & -1.63299 & -0.986771 \\
\hline
$2^{-2}$ & -1.28878 & 2.05237 & 7.0000 & 7 & -13.0639 & -2.82527 \\
\hline
$2^{-1}$ & -3.48234 & 4.29513 & 7.0000 & 7 & -104.512 & -3.53701 \\
\hline
$1$      & -6.86634 & 7.06840 & 7.0000 & 7 & -836.092 & -3.25661 \\
\hline
\end{tabular}
\end{table}
\addtolength{\topmargin}{-1.50pc} \addtolength{\textheight}{3.0pc}
\section{Final remarks}
\label{sec:conclusions} Using the third-order perturbation method  we derived explicit formulae
for the late-time tail (the decay rate \emph{and} the
 amplitude) of a spherically symmetric, self-gravitating massless scalar field for solutions starting
  from small
 initial data. We verified that these formulae are in excellent agreement with the results of numerical
  integration of the
  Einstein-scalar field equations in four and six dimensions. Our results
show that the tail has genuinely \textit{nonlinear} origin and
 should not be mistaken with the linear tail coming from the backscattering off
 the Schwarzschild potential. It seems to us that this distinction between linear and nonlinear
 tails has not been widely recognized in the past which is probably due to the fact that in
 four-dimensional spacetimes
 these two different tails decay at the same rate $t^{-3}$. To
 demonstrate that this coincidence is an idiosyncrasy of four dimensions, we computed both kinds
 of tails in $d+1$ dimensions for $d=2\ell+3$ ($\ell=1,2,...$) and showed that the linear and nonlinear
 tails decay at different rates: $t^{-(6\ell+4)}$ and $t^{-(3\ell+3)}$, respectively. This illustrates
 how viewing the dimension of a spacetime as a parameter may help understand which
 features of general relativity depend crucially on our world being four dimensional and which
 ones are general.

 It would be interesting to generalize the results of this paper to collapsing solutions
 where the endstate of evolution is a black hole. The studies in this direction are in progress
 and will be reported elsewhere. We expect that in this case the tail has
 both linear and
   nonlinear contributions with the latter being qualitatively the same as for dispersive
   solutions described above. Note that the analogous perturbative calculation of tails is
much harder on the black hole background because two basic tools that we used above, Huygens'
principle and the explicit expression for Duhamel's formula, are missing. For dispersive
solutions these tools allowed us to compute the third-order perturbation in a shamelessly
explicit way, however, from the perspective of generalizing the results to collapsing solutions,
it is instructive to redo this calculation in an asymptotic manner keeping track of only leading
order terms in the perturbative equations. Such an efficient calculation, which gives additional
insight into the mechanism of some cancelations in our asymptotic expansions, has been done
recently by Szpak \cite{szpak} and will appear as a comment to this paper.
\vskip 0.2cm \noindent \textbf{Acknowledgments:} We wish to thank Nikodem Szpak for valuable
discussions. We acknowledge support by the MNII grants: NN202 079235 and 189/6.PRUE/2007/7. PB is
grateful to Albert Einstein Institute in Golm and Mittag-Leffler Institute in Djursholm for
hospitality during part of work on this paper.

\end{document}